# Spectral Kinetic Modeling and Long-term Behavior Assessment of *Arthrospira platensis* Growth in Photobioreactor under Red (620 nm) Light Illumination

*Bérangère FARGES\*, Céline LAROCHE, Jean-François CORNET and Claude-Gilles DUSSAP*

Clermont Université - Laboratoire de Génie Chimique et Biochimique, Bât. Polytech.

24, avenue des Landais – BP 206

63174 AUBIERE Cedex - FRANCE

Phone: (33) *4 73 40 55 21

Fax: (33) *4 73 40 78 29

email: Berangere.farges@polytech.univ-bpclermont.fr

**ABSTRACT**

The ability to cultivate the cyanobacterium *Arhtrospira platensis* in artificially lightened photobioreactors using high energetic efficiency (quasi-monochromatic) red LED was investigated. In order to reach the same maximal productivities as with the polychromatic lightening control conditions (red + blue, $P/2e^- = 1.275$), the need to work with an optimal range of wavelength around 620 nm was first established on batch and continuous cultures. The long-term physiological and kinetic behavior was then verified in a continuous photobioreactor illuminated only with red (620 nm) LED, showing that the maximum productivities can be maintained over 30 residence times with only minor changes in the pigment content of the cells corresponding to a well-known adaptation mechanism of the photosystems, but without any effect on growth and stoichiometry. For both poly and monochromatic incident light inputs, a predictive spectral knowledge model was proposed and validated for the first time, allowing the calculation of the kinetics and stoichiometry observed in any photobioreactor cultivating *A. platensis*, or other cyanobacteria if the parameters were updated. It is shown that the photon flux (with a specified wavelength) must be used instead of light energy flux as a relevant control variable for the growth. The experimental and theoretical results obtained in this study demonstrate that it is possible to save the energy consumed by the lightening device of photobioreactors using red LED, the spectral range of which is defined according to the action spectrum of photosynthesis. This appears to be crucial information for applications in which the energy must be rationalized, as it is the case for life support systems in closed environments like a permanent spatial base or a submarine.

**1- INTRODUCTION AND OBJECTIVES**

The use of artificially lightening controlled photobioreactors (PBR) seems a very promising technology if the production of high valuable compounds from photosynthetic micro-organisms is to be envisaged. PBR are also of major interest as main operation in biological life support systems. Such complex artificial ecosystems as the MELiSSA project of ESA (http://www.esa.int/SPECIALS/Melissa/) aim at recycling $CO_2$ gas, liquid and solid wastes produced by a crew in water, food and oxygen inside a thermodynamically closed system (a lunar or Martian base for example). In these systems, the two critical points are the size (or mass) of unit operations and their energy consumption. This is also the case if a previous sub-system and small scale demonstrator is envisaged for a flight onboard the International Space Station such as the Biorat project (1), in which the atmosphere revitalization of a consumer compartment (two mice) is ensured by a fully controlled PBR artificially lightened. In order to minimize in all cases the electrical power consumption of the PBR, the artificial light sources have to be optimized. This can be done by seeking light sources having the best energetic yield for radiation emission or more rigorously, taking into account that photosynthesis is a photonic process, with the best quantum yield. It is well known indeed that water photolysis and charge separation processes at photosystem II (PS II) require a photon of 680 nm maximum wavelength (a minimum energy corresponding to a gap of 1.8 eV). Consequently, a non-negligible part of all the visible photons of shorter wavelength energy is dissipated as heat while photons of longer wavelength cannot participate in charge separation process, their energy being also released as heat. As a consequence, one of the most powerful light sources is likely to be LED (using a wavelength as close as possible to 680 nm) with today a good quantum yield among all the possible existing technologies (and likely to increase significantly in the future decade). This advantage, which makes it possible to save electrical energy and to avoid problems of heat rejection control is counterbalanced by the necessity to operate with quasi-monochromatic light, but the use of LED requires the knowledge of its long-term effect on any photosynthetic cultivation in PBR.

If the effects of monochromatic light on short-term kinetics of $O_2$ evolution have been indeed investigated for a long time, both on eukaryotic microalgae and on cyanobacteria leading to the so-called action spectrum for photosynthesis (strongly different for these two kinds of micro-organisms), the physiological, kinetic and stoichiometric effects of long-term cultivation in continuous PBR are not well documented today. Despite a large amount of qualitative work on the effect of using monochromatic light sources for photosynthetic micro-organisms cultivation (2-7), very few quantitative results have been obtained on cyanobacteria except for the work of Wang *et al*. (8), who has only considered batch cultivations in Erlenmeyer flasks. There is then an important need for studying both the kinetic and related stoichiometric aspects of the well-defined quasi monochromatic light inputs on controlled PBR, first in batch conditions and finally on long-term continuous cultures, enabling us to confirm the potential interest of this approach.

The aim of this paper is first to determine the best quasi-monochromatic wavelength compatible with the objectives of reaching the highest biomass and $O_2$ productivities (related by the photosynthetic quotient, i.e. the global stoichiometry) together with the lowest energy consumption, and secondly to establish and prove that it is possible to envisage a long-term cultivation in a controlled PBR operating in continuous mode. For this purpose, a theoretical investigation of the problem is first led and a predictive knowledge mathematical model for growth in PBR is developed. This model is based on the previous work of the authors (9-11) and enables coupling a fully-predictive light transfer description in the PBR with a knowledge formulation of photosynthetic kinetic rates and stoichiometry from the definition of a quantum yield. This approach, previously established for polychromatic (white) light illumination of the PBR (like all the models of the specialized literature), is generalized for monochromatic illumination leading to a fully-spectral mathematical model (including spectral kinetics coupling) which is formulated in this paper for the first time. In this study, the cyanobacterium *Arthrospira platensis* was chosen for its high potential interest for the atmosphere regeneration objective for life support systems with high pH of cultivation favoring a good $CO_2$ mass transfer and a low $CO_2$ content in the crew compartment. The result of the theoretical model, based on the action spectrum of photosynthesis for *Arthrospira* has demonstrated that working with red LED around 620 nm of wavelength enables us to obtain the same biomass productivity as polychromatic white light. For this reason, in a first experimental section, a red (620 nm) LED panel of high quantum efficiency has been built and its ability to reach (at the same incident photon flux density on the PBR) the same kinetic productivity in batch and continuous cultures as a control polychromatic (red + blue) LED panel was validated. Furthermore, in a second experimental section, the long term (30 residence times) cultivation in continuous PBR using the red (620 nm) LED panel was investigated and performed at a higher incident photon flux density. In both cases, the results obtained were compared with the theoretical predictive model independently developed in this work. Because this model establishes a theoretical link between the stoichiometry observed, the $P/2e^-$ ratio for photosynthesis and the quantum yield (i.e. the kinetic aspects) with the radiant light transfer (the physical limitation by light) and the spectral quality of light, special attention has also been paid on the possible physiological regulation mechanisms of the photosynthetic apparatus which might result and the main photosynthetic pigment contents have been assessed during the long-term continuous experiment.

## 2- MATERIALS AND METHODS

### 2.1- Organism and culture medium

The cyanobacterium *A. platensis* PCC 8005 (Institut Pasteur, Paris, France) was axenically grown in Zarrouk medium, modified by Cogne *et al.* (12), as follow : (g.L$^{-1}$) : NaCl 1.0; CaCl$_2$ 0.03; K$_2$SO$_4$ 1.0; MgSO$_4$, 7 H$_2$O 0.2; K$_2$HPO$_4$ 0.5; NaNO$_3$ 2.5, NaHCO$_3$ 10.5; Na$_2$CO$_3$ 7.6; EDTA 0.08; FeSO$_4$, 7 H$_2$O 0.01 and 1 mL per Liter of medium of trace elements solution (g.L$^{-1}$) : MnCl$_2$, 4 H$_2$O 0.23; ZnSO$_4$, 7 H$_2$O 0.11; CuSO$_4$, 5 H$_2$O 0.03. The final pH of the medium was adjusted at 9.5.

### 2.2- Lighting source

The objectives of this work being to demonstrate that it is possible to cultivate cyanobacteria (mainly *A. platensis*) using monochromatic light (in the range 620-630 nm, see modelling part 3 and the spectral efficiency curve) with the same efficiency as polychromatic white light, we assembled two different LED panels. The first one was conceived as a control panel using polychromatic blue and red LED (the green radiation being poorly absorbed by photosynthetic pigments), in order to rigorously match the photosystem functioning. For this purpose, we chose red and blue LED with the same quantum yields (see Table 1 and Appendix), the number of each type being adjusted to obtain a P/2e$^-$ ratio of roughly 1.3 (see Table 2, stoichiometric equation of part 3 and Appendix) corresponding to the mean observed stoichiometry for incident photon flux densities considered in this study. The second one was clearly conceived as an assay panel, using high efficiency red LED in the correct range of wavelength (see modelling part 3 and the spectral efficiency curve) with a peak emission at 626 nm (see Table 1). For the same electrical power consumption, these LED have been demonstrated (see Appendix) to have a quantum yield 4 times higher than the previous one, delivering therefore an incident photon flux density four times higher (see Table 2) or saving a factor four in electrical power consumption for the same photon flux density.

**Table 1.** Electrical and optical characteristics of the 3 different LED used in the experiments. Condition $T_A$ = 25°C - $I_F$ = 20mA

|  | **Agilent HLMP-EG08-WZ000 Red LED** | **Toyoda Gosei E1L51-3B0A Blue LED** | **Fairchild MV8014 Red LED** |
|---|---|---|---|
| **Luminous intensity (mcd)** | 16000 | 1800 | 1500 |
| **Intensity (mA)** | 20 | 20 | 20 |
| **Typical voltage (V)** | 1.9 | 3.4 | 2.1 |
| **Peak wavelength (nm)** | 626 | 470 | 640 |
| **Viewing angle (°)** | 8 | 15 | 12 |

The two LED panels (180 mm height × 105 mm width) were built in the same manner, using 180 LED mounted with 10 mm between them.

**Table 2.** Maximal incident photon flux density for the two LED panels used in this study. The polychromatic control panel is conceived with a $P/2e^-$ = 1.275 blue LED for one red LED.

|  | LED | Wavelength (nm) | Maximal incident photon flux density (µmol.m$^{-2}$.s$^{-1}$) |
|---|---|---|---|
| High efficiency red LED panel | Red | 626 | 135 |
| Polychromatic control (red + blue) LED panel | Red | 640 | 14.5 |
|  | Blue | 470 | 18.5 (= 14.5 × 1.275) |
|  | Total | 470 and 640 | 33 |

The maximal incident photon flux densities $q_\cap$ (in µmol.m$^{-2}$.s$^{-1}$) delivered by the two panels were determined by averaging 24 points of measurement on the surface and using a LI-COR quantum sensor (LI-190 SA). The results are summarized in Table 2 and show, as theoretically calculated (see

Appendix), that the high efficiency red LED panel reaches a 4-times higher photon flux density for the same electrical power than the control panel. Conversely, in order to obtain the same incident photon flux densities for the previous experiments this panel was used with a 25% electrical power consumption, delivering then 33 $\mu mol.m^{-2}.s^{-1}$ for the two panels.

### 2.3- Experimental device

A scheme and a picture of the experimental device are presented in Figure 1. Experiments in batch were performed using two identical 600 mL working volume rectangular photobioreactors with different LED panels in parallel. The first panel corresponding to polychromatic (red + blue) light served as control panel and a second panel with high efficiency monochromatic red LED, corresponding to the main purpose of this work, was also assembled and used. A black plate was placed between the two LED panels in order to avoid interferences between the polychromatic and the monochromatic light. In order to maintain an optimal temperature of 36°C in the PBR, a temperature sensor was used as external control for a heating bath, with water circulation in heating jackets. The distance between LED and PBR was fixed at 10 cm, due to magnetic stirring devices used to mix cultures homogeneously (see Figure 1b). The cultures were aerated with ~2% $CO_2$ enriched air for pH = 9.5 control, supplied to the PBR using 2 mass flow controllers for gas mixing. In order to prevent liquid losses, peltier condensers were used.

For experiments in continuous, the same experimental device supplemented with a peristaltic pump was used. Two flexible tubes, one for the input and one for the output have been fitted inside this circular pump in order to obtain equal input and output flows.

During the experiments, samples were taken twice a day and pH, $OD_{750nm}$, dry mass and pigment content were analysed.

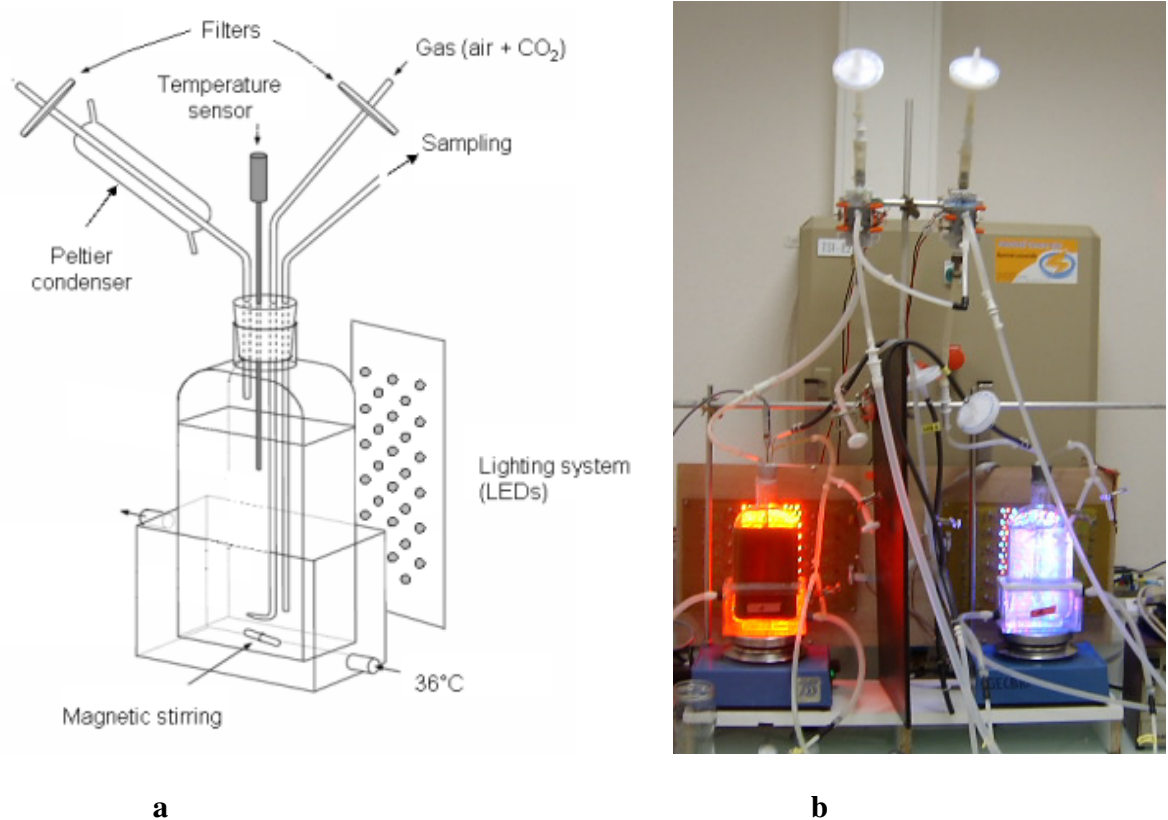

         **a**                  **b**

**Figure 1.**   a) Scheme of the experimental device for rectangular PBR (600 mL working volume) experiments with LED panels.

b) Picture of the experimental device for rectangular PBR experiments with the control polychromatic (red + blue) LED panel (right) and the high efficiency monochromatic red LED panel (left).

### 2.4- Experimental methods

*Biomass determination*

The biomass concentration was determined turbidimetrically at 750 nm (spectrophotometer SAFAS, MC2) using a conversion factor for dry weight and confirmed by dry weight measurements (filters are dried in an oven at 110°C during 24 hours).

*Pigment assay*

Pigment assay has been performed by measuring optical density at 620, 678, and 750 nm after sonication (to break *A. platensis* trichomes). Using correlations and correction factors previously established (13, 15), it was possible to quantify (in g.L$^{-1}$) chlorophyll *a* and phycocyanin pigments as follows:

[chl *a*] = $1.04 \times 10^{-2}$ A$_{678}$ – $4.09 \times 10^{-3}$ A$_{620}$

[phyc] = 0.297 A$_{620}$ – 0.076 A$_{678}$

in witch A$_{620}$ = OD$_{620}$ - 0.93 OD$_{750}$ and A$_{678}$ = OD$_{678}$ - 0.89 OD$_{750}$

*Action spectrum determination*

The *A. platensis* action spectrum, corresponding to the measurement of O$_2$ evolution rate for a given monochromatic incident radiation was performed using an ultra-fast Haxo-Blinks type O$_2$ electrode, associated to a (xenon lamp) light source mounted with a monochromator in series (13).

## 3- SPECTRAL KINETIC KNOWLEDGE MODEL FOR *Arthrospira platensis* GROWTH

The best way to achieve a structured knowledge model of photobioreactors which had been previously established by the authors (10), requires a precise formulation of the photon transport in a PBR of a given geometry and boundary conditions, and then the coupling between the so-called local volumetric rate of the radiant energy absorbed (LVREA) and the local kinetic rates from the knowledge of energetic and quantum yields. Finally, the complete model is obtained by averaging rates over the illuminated volume of the reactor and including them in the well-known elemental (dry weight) biomass balance to give the evolution equation for any considered component, if a stoichiometric equation is used. The extension of this approach, initially developed in the photosynthetically active radiation (PAR) domain (10, 16) to a spectral model, requires first to have a predictive method in calculating the spectral radiative properties of any micro-organism in order to solve accurately the spectral radiative transfer equation (RTE). Secondly, it requires knowing the effect of using a narrow wavelength bandwidth on kinetic rates, i.e. the so-called action spectrum for photosynthesis.

### 3.1- Spectral Light Transfer Model

Some of the authors have already demonstrated how it was possible to obtain the spectral radiative properties of *Chlamydomonas reinhardtii* by a predictive mean, using an original method enabling first the calculation of the spectral optical properties of photosynthetic micro-organisms (the complex refractive index) by convolution, from data banks of *in vivo* absorption spectrum of pigments (11). In this paper, the same approach has been used for *Arthrospira platensis*, but the considered spectral radiative properties (mass absorption $Ea_\lambda$ and mass scattering $Es_\lambda$ coefficients with the phase function $p_\lambda$) have been calculated from the predictive Lorenz-Mie theory, considering *A. platensis* as a long circular cylinder (17) randomly oriented (i.e. averaging over all the incident directions of light), and the anomalous diffraction approximation of Van de Hulst (18). Finally, the RTE is solved in rectangular PBR illuminated on one side, using a one-dimensional approximation and a normal quasi-collimated two-flux generalized model, taking into account the reflection at the rear (thickness $L$) by the water-glass interface (say $r_\lambda = 0.01$ from the geometrical optics calculation). In this case, the irradiance profile is given by (11):

$$\frac{G_\lambda(z)}{q_{\lambda,\cap}} = \frac{[r_\lambda(1+\alpha_\lambda)\exp(-\delta_\lambda L)-(1-\alpha_\lambda)\exp(-\delta_\lambda L)]\exp(\delta_\lambda z)+[(1+\alpha_\lambda)\exp(\delta_\lambda L)-r_\lambda(1-\alpha_\lambda)\exp(-\delta_\lambda L)]\exp(-\delta_\lambda z)}{(1+\alpha_\lambda)^2\exp(\delta_\lambda L)-(1-\alpha_\lambda)^2\exp(-\delta_\lambda L)-r_\lambda(1-\alpha_\lambda^2)\exp(\delta_\lambda L)+r_\lambda(1-\alpha_\lambda^2)\exp(-\delta_\lambda L)} \quad [1]$$

in which we have introduced the linear scattering modulus:

$$\alpha_\lambda = \sqrt{\frac{Ea_\lambda}{Ea_\lambda + 2b_\lambda Es_\lambda}} \quad [2]$$

and the extinction coefficient related to the dry biomass concentration $C_X$ by:

$$\delta_\lambda = C_X \sqrt{Ea_\lambda(Ea_\lambda + 2b_\lambda Es_\lambda)} \quad [3]$$

In these two definitions, the back-scattered fraction $b_\lambda$ characterizing the phase function is calculated as:

$$b_\lambda = \frac{1}{2} \int_{\pi/2}^{\pi} p_\lambda(\theta,\theta') \sin\theta \, d\theta \quad [4]$$

Clearly, the set of equations [1-4] is entirely predictive, and avoid using models of representation (19-20) in the light transfer description. At this stage, it should be emphasized that the previous irradiance $G$ and incident flux density $q_\cap$ must be defined as photonic quantities (in $\mu mol.m^{-2}.s^{-1}$), making it possible to formulate the kinetic coupling solely in terms of quantum yield.

### 3.2- Coupling Spectral Light Transfer and Local Kinetic Rates

As explained elsewhere (10), the local coupling demands to know the LVREA deriving from the previous irradiance profile (eq. 1):

$$\mathcal{A}_\lambda(z) = Ea_\lambda C_X G_\lambda(z) \quad [5]$$

and indeed the conversion efficiency of this absorbed photons rate. From first principles analysis of excitation energy transfer in antenna and of the so-called Z-scheme for photosynthesis, it is possible to introduce a local energetic yield of conversion, depending on the local wavelength-averaged irradiance (dissipative part of the photonic energy as a non-linear phenomenon):

$$\rho = \rho_M \frac{K}{K+G} \quad [6]$$

and a conservative mole quantum yield for the Z-scheme $\phi'$ which is only P/2e$^-$ dependent (10). If an overall structured stoichiometry is postulated, the associated Z-scheme stoichiometry for photosynthesis leads easily to the following definition for $\phi'$ (see Figure 2):

$$\phi' = \frac{1}{2\upsilon_{NADPH,H^+-X}(1+P/2e^-)} \quad [7]$$

As long as polychromatic radiation was used, this formula does not depend on the spectral nature of the photons ($\lambda$), but only on the stoichiometry for the produced biomass which is affected only by the averaged radiation field (10). For reasonably low incident light fluxes (lower than 150 µmol.m$^{-2}$.s$^{-1}$), the following structured stoichiometry has been proposed for *A. platensis* (9):

$$CO_2 + 1.463\,H_2O + 0.173\,HNO_3 + 0.006\,H_2SO_4 + 3.544\,ATP + 2.779\,(NADPH,H^+)$$
$$\xrightarrow{J_X} CH_{1.575}O_{0.459}N_{0.173}S_{0.006}P_{0.006} + 3.538\,P_i + 3.544\,ADP + 2.779\,NADP^+ \quad [8]$$

from which the associated Z-scheme stoichiometry for photosynthesis is straightforward:

$$2.779\,NADP^+ + 2.779\,H_2O \xrightarrow{J_{COF}} 2.779\,(NADPH,H^+) + 1.389\,O_2$$
$$3.544\,(ADP+P_i) \xrightarrow{J_{ATP}} 3.544\,ATP + 3.544\,H_2O \quad [9]$$

Obviously, in this case, the mean value of the P/2e$^-$ ratio is:

$$P/2e^- = \frac{J_{ATP}}{J_{COF}} = 1.275 \quad [10]$$

leading from (eq. 7, 9) to $\phi' = 7.9 \times 10^{-8}$ mol$_X$ / µmol$_{h\nu}$. Because mass growth rates are rather considered, the mass quantum yield $\phi$ is easily derived from the corresponding C-molar mass for *A. platensis* $M_X = 23.7 \times 10^{-3}$ kg$_X$.mol$_X^{-1}$, giving:

$$\phi = M_X \phi' = 1.85 \times 10^{-9} \text{ kg}_X / \text{µmol}_{h\nu}. \quad [11]$$

In the same way, the local yield $\rho$ is weakly dependent on the wavelength and may be considered as constant over the PAR. Its maximum value $\rho_M = 0.8$ for a quasi-collimated field has been calculated elsewhere (21), and the half saturation constant $K = 90$ µmol.m$^{-2}$.s$^{-1}$ has been validated for *A. platensis* over a wide range of operating conditions and reactor geometries (9, 16, 22, 23). Finally, using a polychromatic irradiation (i.e. visible light) for the PBR leads to the following local kinetic rates (with the help of eq. 5 and 6) for biomass concentration (10, 16):

$$r_X = \rho_M \phi \frac{K}{K+G} \mathcal{A} = \rho_M \phi K Ea \frac{G}{K+G} C_X \quad [12]$$

Of course, in this equation, $Ea$, $G$ and $\mathcal{A}$ are mean spectral quantities on the visible spectrum which can be obtained rigorously by averaging spectral LVREA (11), or working directly with averaged radiative properties if one accepts the corresponding lack of accuracy (11).

Nevertheless, using quasi-monochromatic light for the cultivation of photosynthetic micro-organisms leads to different kinetic behaviors depending on how the functioning of the two photosystems in series is affected. For eukaryotic micro-organisms, the action spectrum ranges between 400 and 680 nm with the same efficiency for photosynthesis whatever the wavelength in the interval; then it is well known that it is possible to use red LED (for example) with the same efficiency as white light (3). Conversely, for cyanobacteria like *A. platensis*, the PSII antennas are mainly constituted of phycocyanins and allophycocyanins (phycobilins) absorbing at their maximum around 620 nm, whereas the PSI is mainly constituted by chlorophylls (absorbing at 440 and 680 nm, see Figure 2).

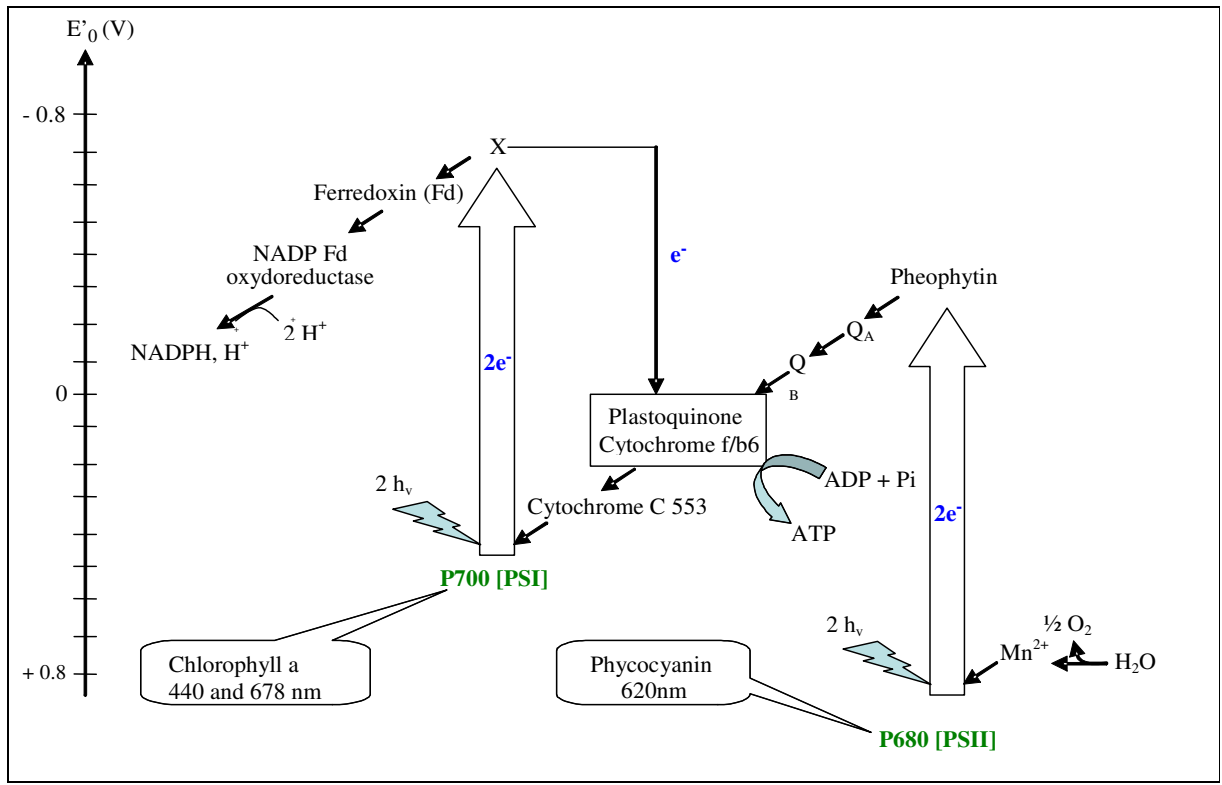

**Figure 2**. Typical Z-scheme of photosynthesis for cyanobacteria (from 9)

If quasi-monochromatic light is used, only red photons can efficiently perform water photolysis at PSII. On the contrary, blue photons cannot be trapped easily by PSII, but only by PSI, leading to a poor efficiency of photosynthesis in this case, resulting from the fact that the two photosystems have to work in series. If only red photons are used, a well-established compensation phenomenon exists (24-26), enabling the redistribution of the electrons flux to the PSI with a high efficiency and equilibrating the P/2e$^-$ ratio. These phenomena are evidenced on the so-called action spectrum of photosynthesis, obtained from the measurement of $O_2$ evolution rates using monochromatic light (13, 14). Such experiment is depicted in Figure 3 for *A. platensis*. This action spectrum enables the introduction of the spectral information in the kinetic coupling by defining a spectral efficiency $e_\lambda$ (Figure 3) reaching its maximum value of 1 (as in polychromatic white light) in the range 618-626 nm (with less than 1% deviation). This figure clearly demonstrates that it is necessary to choose LED in this range (around 620 nm) in order to cultivate cyanobacteria (particularly *A. platensis*) with the same efficiency as with polychromatic visible light.

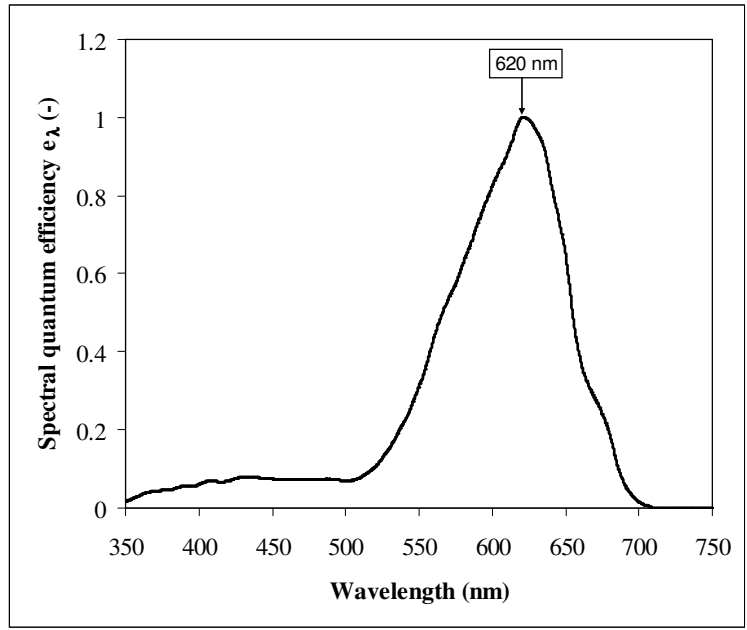

**Figure 3**. Determination of the spectral efficiency $e_\lambda$ for *A. platensis* from the experimental action spectrum of photosynthesis in the PAR measured with Haxo-Blinks electrode (13, 14).

Actually, using LED as lightening source is never really monochromatic and all the spectral quantities need to be averaged on a small bandwidth (usually 20 nm) as follows (the final accuracy is not affected in this case taking into account the narrow bandwidth of the considered spectrum):

$$Ea_{\Delta\lambda}, Es_{\Delta\lambda}, b_{\Delta\lambda}, e_{\Delta\lambda} = \frac{1}{\Delta\lambda} \int_{\lambda_{min}}^{\lambda_{max}} (Ea_\lambda, Es_\lambda, b_\lambda, e_\lambda) d\lambda \quad [13]$$

$$G, q_\cap = \int_{\lambda_{min}}^{\lambda_{max}} (G_\lambda, q_{\lambda,\cap}) d\lambda \quad [14]$$

The values of the radiative properties for the wavelength bandwidth considered in this study and calculated as explained above are summarized in Table 3 (respectively for 470, 626, and 640 nm $\pm 2\sigma = \pm 10$ nm). Introducing then these definitions in equation [12], together with the knowledge of the spectral efficiency $e_\lambda$ (Fig. 3), leads to a generalization of the kinetic coupling for spectral applications as:

$$r_X = e_{\Delta\lambda}\, \rho_M\, \phi\, K\, Ea_{\Delta\lambda} \frac{G}{K+G} C_X \quad [15]$$

in which the averaged irradiance $G$ and the averaged incident light flux $q_\cap$ are always defined by mean of equation [14].

**Table 3.** Main radiative properties versus the considered wavelength bandwidth for the LED of the study, as predictive data for light transfer model calculation and obtained from the method developed by Pottier *et al.* (11), for the optical properties assessment. These values were obtained applying the Lorenz-Mie theory with the anomalous diffraction approximation, assimilating *Arthrospira platensis* to a long circular cylinder randomly oriented.

| Wavelength bandwidth $\Delta\lambda = \bar{\lambda} \pm 10$ nm | $\bar{\lambda}$ = 470 nm (type Toyoda Gosei E1L51-3B0A) | $\bar{\lambda}$ = 626 nm (type Agilent HLMP-EG08-WZ000) | $\bar{\lambda}$ = 640 nm (type Fairchild MV8014) |
|---|---|---|---|
| **Imaginary part of the refractive index $\bar{\kappa}$ (dimensionless)** | $2.34 \times 10^{-3}$ | $4.47 \times 10^{-3}$ | $3.17 \times 10^{-3}$ |
| **Average mass absorption coefficient $Ea_{\Delta\lambda}$ (m$^2$.kg$^{-1}$)** | 225 | 225 | 201 |
| **Average mass scattering coefficient $Es_{\Delta\lambda}$ (m$^2$.kg$^{-1}$)** | 606 | 609 | 677 |
| **Average backscattered fraction $b_{\Delta\lambda}$ (dimensionless)** | 0.040 | 0.033 | 0.031 |
| **Average linear scattering modulus $\alpha_{\Delta\lambda}$ (dimensionless)** | 0.907 | 0.921 | 0.911 |

### 3.3- Mean Spatial Kinetic Rates and Evolution Equations for the PBR

As already explained in previous papers (9, 10, 16, 22, 23), the average volumetric biomass growth rate in the PBR is then obtained by averaging local rates onto the active metabolic and illuminated volume fraction in the reactor (assuming that the dark part of the PBR has no metabolic contribution as it is the case for cyanobacteria having their respiratory electron transfer chain inhibited by light). If this fraction $\gamma$ is defined as the part of the volume ($V_l$) in which the irradiance is higher than the value of the compensation point for photosynthesis, the global productivity is calculated and further simplified, taking into account the one-dimensional geometrical simplification:

$$<r_X> = \gamma \frac{1}{V_l} \iiint_{V_l} r_X \, dV = \gamma \frac{1}{L_l} \int_0^{L_l} r_X \, dz \quad [16]$$

In case of kinetic regime functioning, photons are in excess in the reactor, $\gamma = 1$ and $L_l = L$, leading to an integration over the entire volume of the PBR. The evolution equation for the biomass concentration is then easily established, applying a mass balance on the perfectly mixed homogeneous PBR and introducing the hydraulic residence time $\tau$, giving:

$$\frac{dC_X}{dt} = <r_X> - \frac{C_X}{\tau} \quad [17]$$

and in continuous and steady state operation:

$$C_X = <r_X> \tau \quad [18]$$

These balances, associated to the stoichiometric equation [8], completed the spectral knowledge model, as a powerful tool in PBR analysis, simulation, sizing and model based predictive control (10, 16).

## 4- RESULTS AND DISCUSSION

As explained in the introduction, the objective of this work was to prove the feasibility of a long term cultivation of *A. platensis*, using high energetic efficiency red LED (instead of white or any polychromatic radiation with lower efficiency) with the same kinetic performance as for visible light. The preceding mathematical modeling section clearly established that it was theoretically possible from the knowledge of the spectral action spectrum efficiency (Figure 3), taking LED with a main range of emission between 618 and 626 nm, leading to our choice of both the control and the red LED panels (see Materials and Methods section and Appendix). Nevertheless, and before discussing the experimental results obtained in this work, it is interesting to compare the previous mathematical model with the data available in the literature.

### 4.1- Comments of some results in the literature

Only very few quantitative results (with well-documented PBR characteristics and accurate incident quantum light flux calibration) exist for monochromatic light cultivation of *A. platensis* (or other cyanobacteria) in the literature. We can already discuss further the results of Wang *et al.* (8), who studied, the effects of different LED on kinetics of *Spirulina platensis* in flasks. Considering that their more accurate results were obtained at high light intensity for the LED (3000 $\mu mol.m^{-2}.s^{-1}$) and assuming that they worked in physical limitation by light (linear biomass concentration time course; constant growth rate), it is possible to calculate the best observed productivity (a control), and then the relative productivity for other wavelength sources, matching closely the definition of the spectral efficiency $e_\lambda$ (Figure 3 and eq. 15). Fortunately, their higher productivity was observed for a red LED emitting in the range 620-645 nm corresponding roughly to the highest value of the spectral efficiency ($e_{\Delta\lambda} = 1$) we have introduced (Figure 3). In the same way, calculating their values of $e_{\Delta\lambda}$ for green LED (515-540 nm) and blue LED (460-475 nm) relative to the productivity with red LED (control), we get respectively 0.25 ± 0.04 and 0.08 ± 0.01, whereas our theoretical results (Figure 3) for $e_{\Delta\lambda}$ would lead to 0.22 and 0.074. This preliminary analysis shows that our theoretical definition of the spectral efficiency tallies with the experimental results of Wang *et al*. (8). On the contrary, the agreement was not correct for yellow LED (587-595 nm) giving roughly the same value of $e_{\Delta\lambda}$ as the green LED for experimental productivity, whereas our theoretical value would be closer to 0.75. This surprising discrepancy could result in some experimental details that have not been reported in their paper like for example a slight difference in the incident intensity or in the flask geometry; this could also explain the slightly lower productivity observed for white light (the control) in comparison to the red LED productivity (apparently not possible from a theoretical point of view).

In any case, it clearly appears that a complete validation of the spectral efficiency curve on kinetic growth rates (eq. 15-17) would require further investigations in fully controlled PBR and many different spectral incident light inputs, but this remains out of the scope of this paper.

**4.2- Short-term comparison between control LED panel and high efficiency red LED panel (validation of the knowledge model)**

The previous theoretical analysis and mathematical modeling evidence that cultivating *A. platensis* in two identical PBR with the same incident photon flux density $q_\cap$, using a polychromatic control (red + blue, P/2e$^-$ = 1.275) LED panel or a high efficiency monochromatic red LED panel (with $e_{\Delta\lambda} = 1$) should lead to the same kinetic growth rates, if operating in physical limitation by light (all the incident photons absorbed). Consequently, a first experiment was carried out in batch conditions, taking the incident photon flux density at $q_\cap = 33$ μmol.m$^{-2}$.s$^{-1}$ (the maximum value for the control LED panel, see Table 2 and Appendix) for the two PBR, and with a concentration inoculum sufficiently high to obtain the physical limitation by light as confirmed by the predictive light transfer model (eq. 1-4). As expected, the resulting biomass concentration time course observed in Figure 4 appears as linear, the slope being the average biomass productivity (22). The results depicted in Figure 4 correspond to two-fold repeated independent experiments in the two PBR, leading respectively to a mean productivity of $<r_X> = (3.2 \pm 0.3) \times 10^{-3}$ g.L$^{-1}$.h$^{-1}$ for the control LED panel and $<r_X> = (3.3 \pm 0.4) \times 10^{-3}$ g.L$^{-1}$.h$^{-1}$ for the red LED panel, demonstrating that, as foreseen, there is no significant kinetic difference between the two kinds of LED panel, and confirming then qualitatively the previous reasoning.

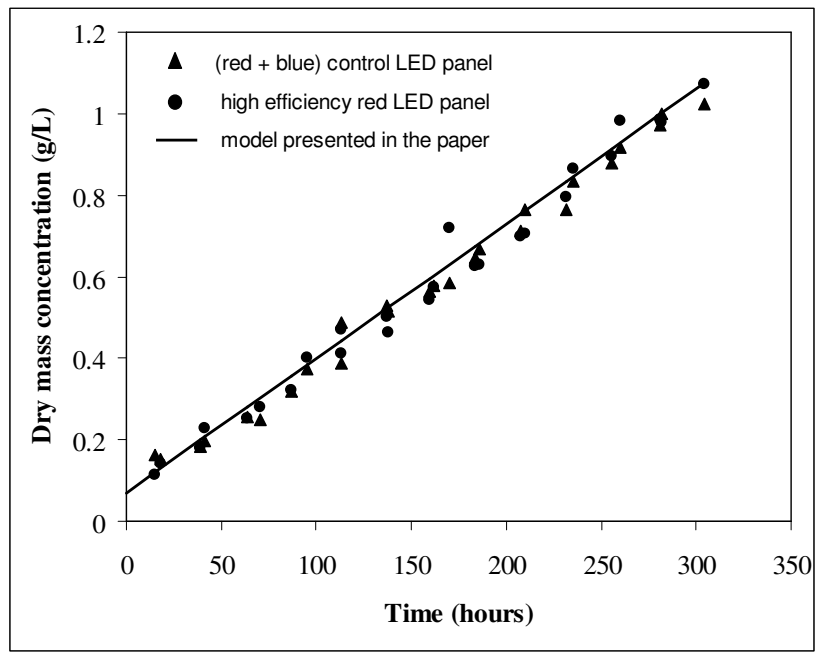

**Figure 4**. Time course of *Arthrospira platensis* dry mass concentration for batch cultivation in two identical rectangular PBR illuminated on one side ($L$ = 0.04 m), with a control LED panel and a high efficiency LED panel delivering an incident photon flux density $q_\cap$ = 33 µmol.m$^{-2}$.s$^{-1}$ (corresponding to 25% of the maximum output for the latter as explained in appendix); comparison between the completely predictive model and the experimental results.

Moreover, as confirmed by the continuous line in Figure 4 representing the knowledge model calculation (eq. 1-16 with $e_{\Delta\lambda} = 1$ for monochromatic red LED panel), the experimental productivities are also quantitatively in very good agreement with the theoretical approach proposed in this paper, leading with this incident light flux and this reactor geometry to $<r_X> = 3.3 \times 10^{-3}$ g.L$^{-1}$.h$^{-1}$.

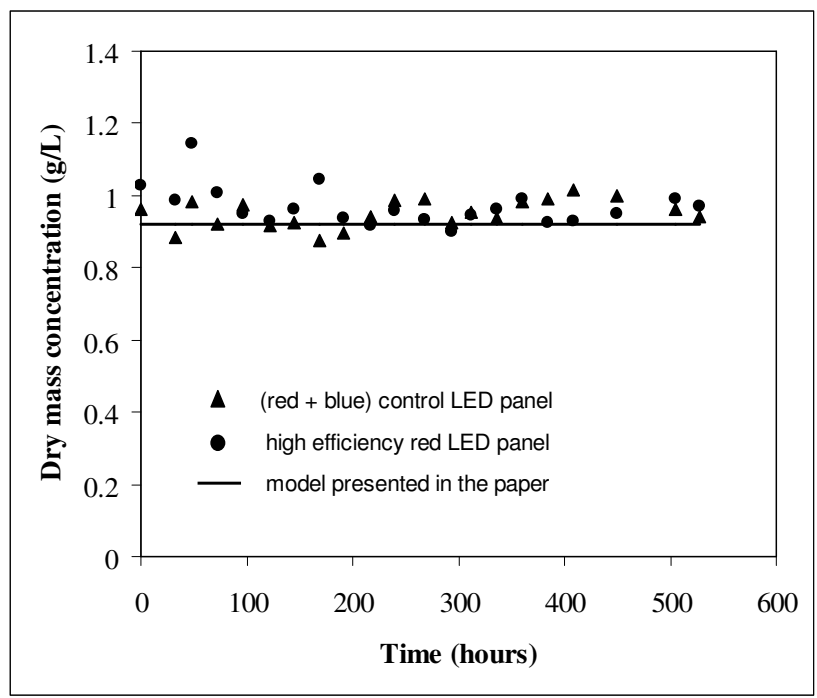

**Figure 5**. *Arthrospira platensis* output dry mass concentration versus time for continuous cultivation in two identical rectangular PBR illuminated on one side ($L = 0.04$ m), with a control LED panel and a high efficiency LED panel delivering an incident photon flux density $q_\cap = 33$ µmol.m$^{-2}$.s$^{-1}$ (corresponding to 25% of the maximum output for this later as explained in appendix); comparison between the completely predictive model and the experimental results. The corresponding residence time was 272 hours.

Besides, these results were also confirmed in continuous cultivation conditions, using again the same incident photon flux density $q_\cap$ = 33 µmol.m$^{-2}$.s$^{-1}$ for the two LED panels and the same hydraulic residence time $\tau$ = 272 hours for the two PBR. The experimental data are reported in Figure 5 and compared to the theoretical value of outlet biomass concentration given by the model (eq. 1-15 and 18). They still demonstrated that first, there is no significant difference between the control LED panel and the red one, and second that they are very close to the theoretical knowledge model calculations.

These results clearly confirm the point of interest in using high efficiency red LED panel for the cultivation of cyanobacteria in PBR, since the same productivity was obtained as with the polychromatic control LED panel but saving a factor 4 for the electrical power consumption in this case (see appendix for detailed calculations). This is clearly a crucial point if the availability in electrical power is strongly constrained as it is the case onboard the International Space Station for experiments testing the microgravity effects.

**4.3- Long-term validation using the high efficiency red LED panel and pigment content modifications**

Finally, in order to verify the long term behavior of *A. platensis* cultures in PBR illuminated with high efficiency red LED panel, a continuous culture was maintained over 3500 hours (5 months), this choice corresponding to the future duration of an experiment onboard the International Space Station. The high efficiency panel (red LED) was used at its nominal electrical power, delivering an incident photon flux density $q_\cap$ = 135 µmol.m$^{-2}$.s$^{-1}$ (see Table 2 and appendix) and the residence time in the PBR was decreased to $\tau$ = 62.5 hours.

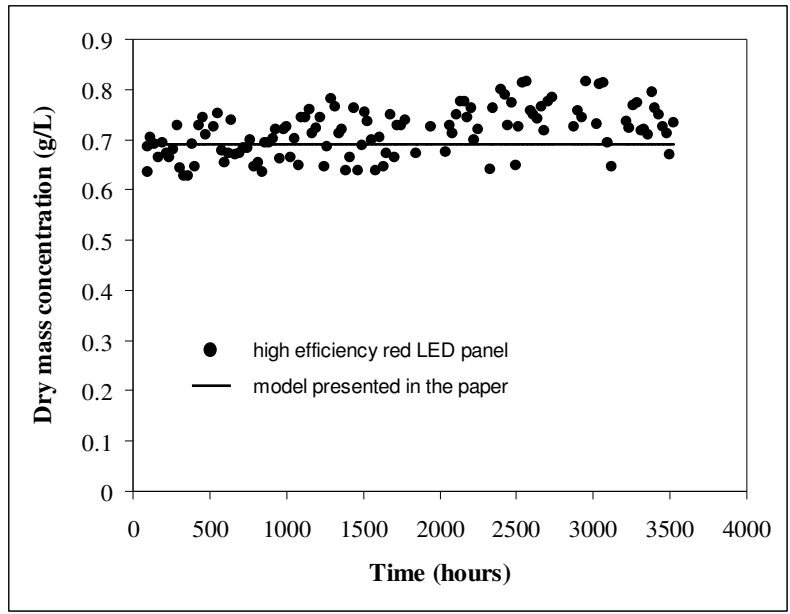

**Figure 6.** *Arthrospira platensis* output dry mass concentration versus time for a long-term continuous cultivation in rectangular PBR illuminated on one side ($L = 0.04$ m) and using a high efficiency LED panel delivering an incident photon flux density $q_\cap = 135$ µmol.m$^{-2}$.s$^{-1}$; comparison between the completely predictive model and the experimental results. The residence time was 62.5 hours.

The experimental data obtained which are summarized in Figure 6 show again a very good agreement with the model calculation (continuous line) leading to a mean productivity in this case $<r_X> = 1.1 \times 10^{-2}$ g.L$^{-1}$.h$^{-1}$ i.e. to a theoretical output biomass concentration (eq. 18) $C_X = 0.7$ g/L. This confirms again the predictive character and the structure of the developed knowledge model with an incident light flux four times higher. Besides, these results also demonstrate that there was no kinetic deviation in the PBR driving a long term cultivation corresponding to 30 residence times for the complete experiment.

In addition, the long-time red light cultivation effect on the pigment contents of the cells was investigated. The results are reported in Table 4, in comparison with the pigment contents also obtained for the (red + blue) control LED panel and the well-known contents obtained for white light (9). As shown, using polychromatic (red + blue) control LED panel does not modify the pigment contents observed for true white light as previously expected, whereas the use of monochromatic red LED panel resulted in a two-fold decrease of the phycocyanin content.

**Table 4.** Mean pigment contents (% DM) obtained for long-time continuous cultivation of *A. platensis* in PBR using different spectral incident light sources and for incident photon flux densities $q_\cap$ lower than 150 $\mu mol.m^{-2}.s^{-1}$.

|  | **White light or any polychromatic light** | **(red + blue) polychromatic control LED panel** | **High efficiency red LED panel** |
|---|---|---|---|
| **Chlorophyll *a* content** | 0.8 ± 0.1 % | 0.8 ± 0.1 % | 0.8 ± 0.1 % |
| **Phycocyanin content** | 16 ± 1 % | 16 ± 1 % | 7.5 ± 0.5 % |

These results were also confirmed by the visible light absorption spectra obtained for the two kinds of LED panels as shown in Figure 7. These spectra were normalized at 750 nm to take into account a possible effect of the biomass concentration on the turbidity of the sample.

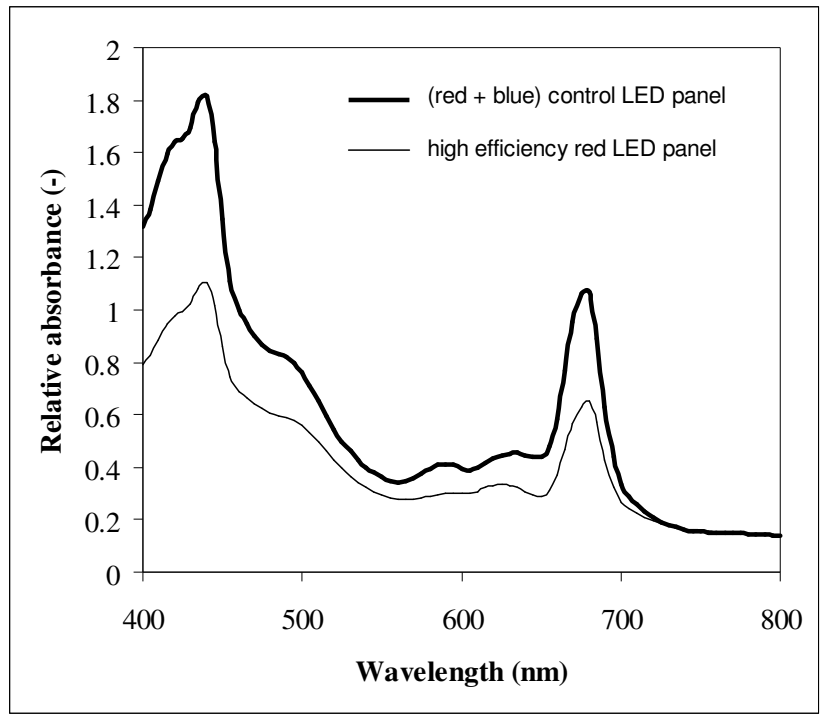

**Figure 7**. Absorption spectra obtained from *A. platensis* samples being cultivated with (red + blue) control LED panel and high efficiency red (620 nm) panel. The spectra are normalized from the turbidity of the sample (dielectric behavior of cells) between 750 and 800 nm.

A possible explanation of these pigment content changes could come from in the well-known compensation mechanism of the electron fluxes in the two photosystems already mentioned in the modeling section and explaining the action spectrum for cyanobacteria. The redirection of the electron flux to the PSI from the mainly excited PSII (red light at 620 nm) seems also responsible in a physiological regulation mechanism decreasing the efficiency of photon capture in the antennas of the PSII (the phycobilins). Conversely, the chlorophyll content, rather representative of the number of reaction centers for photosynthesis, remains constant in all spectral illumination conditions.

It must be pointed out at this stage that this physiological regulation mechanism had likely no significant effect on the mean global stoichiometry in the PBR (mainly on the photosynthetic quotient $Q_P = <r_{O_2}>/<r_{CO_2}>$) because as demonstrated by the mathematical knowledge model (eq. 8-11), the stoichiometry of the produced biomass, the mean P/2e$^-$ ratio and the kinetics (the mean growth rates) are directly interrelated, a modification of any one of these variables affecting obligatorily the other.

### 4.4- Recovery in pigment content

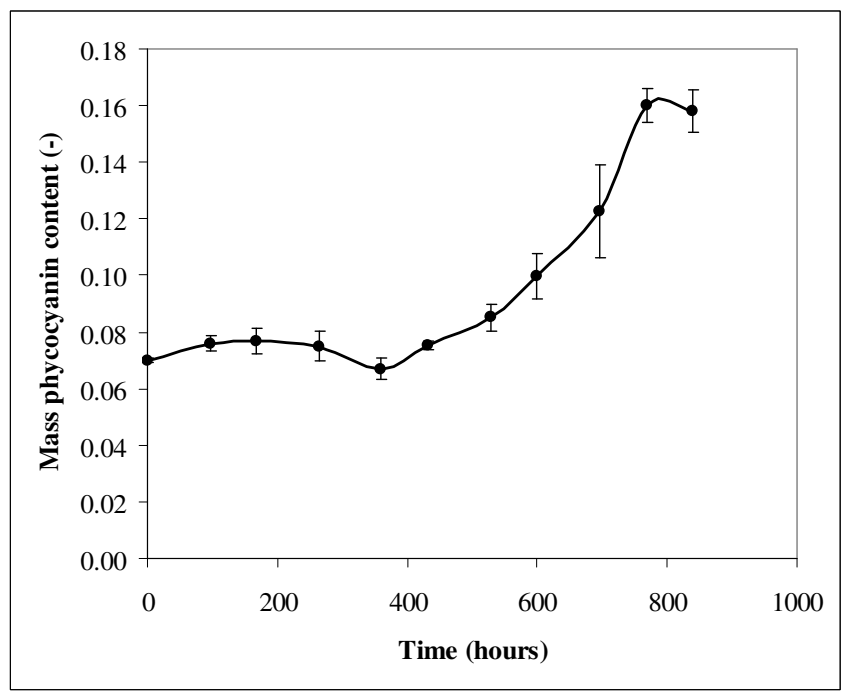

**Figure 8**. Time course recovery of phycocyanin pigment content during the cultivation (for *A. platensis*) in standard conditions (red + blue control LED panel) after long-term continuous exposure to red (620 nm) light. The corresponding residence time of cells in the continuous PBR was 272 hours and the incident photon flux density $q_\cap = 33$ µmol.m$^{-2}$.s$^{-1}$. The initial values of pigment content are obtained from continuous culture with red LED, $q_\cap = 135$ µmol.m$^{-2}$.s$^{-1}$ and $\tau = 62.5$ h.

At the end of the previous experiment, the cells culture conditions were changed, applying on the reactor again a (red + blue) control LED panel with an incident photon flux density $q_\cap = 33$ µmol.m$^{-2}$.s$^{-1}$ and a dilution rate corresponding to a hydraulic residence time of 272h, enabling us to study the conditions for pigment content recovery. As depicted in Figure 8, the total recovery in pigment content for phycocyanin takes about 850 hours. This time finally corresponds to three residence times in the reactor which is a well-known characteristic time for most of the continuous processes.

## 5- CONCLUSIONS

In this paper, the possibility of saving significant electrical energy power consumption in artificially lightened PBR was demonstrated, using a high efficiency red (620 nm) LED panel instead of classical white sources or any other polychromatic sources. The choice of the most efficient quasi-monochromatic wavelength range (618-626 nm) for *Arthrospira platensis* was theoretically established from the action spectrum of photosynthesis, and it has been proved experimentally that the resulting biomass productivity $<r_X>$ was as high as that obtained with a polychromatic (red + blue) control LED panel, especially designed to match the P/2e$^-$ ratio of *A. platensis* at low incident photon flux densities (less than 150 µmol.m$^{-2}$.s$^{-1}$). The comparison was first led in two identical PBR lightened with the same low incident photon flux density (33 µmol.m$^{-2}$.s$^{-1}$) for the two LED panels (control and assay) and operating in batch and then in continuous mode. The long term behavior of the continuous cultivation of *A. platensis* was investigated at a higher incident photon flux density (135 µmol.m$^{-2}$.s$^{-1}$) delivered by the high efficiency red (620 nm) LED panel, with the same energy consumption. The results demonstrated the possibility of obtaining high and stable biomass productivities (over 30 residence times, i.e. 3500 hours) with only minor physiological modifications as the decrease in phycocyanin content in the cells, corresponding to a well-known electron fluxes equilibration mechanism in the Z-scheme for photosynthesis.

Additionally, a predictive kinetic and stoichiometric spectral knowledge model was established in the paper for the first time, introducing the spectral efficiency $e_\lambda$ in coupling kinetics with the volumetric light absorption rate. The resulting simulations were proved to be in very good agreement with the experimental data, both for classical polychromatic irradiation but also when using the monochromatic red LED panel conceived on its theoretical basis. Because all the parameters of the model can be calculated by a predictive mean (optical and radiative properties, quantum yield), its structure appears as very general and it could be used for any other micro-organism as long as the action spectrum for photosynthesis and a structured stoichiometry is known. It must be emphasized that the proposed model clearly established the link between global stoichiometry, $P/2e^-$ ratio (thermodynamics of photosynthesis), and the quantum yield responsible for the kinetic aspects for biomass, substrates and products. Conversely, the fact that the productivities appeared insensitive to the spectral nature of light when using the optimal red range of wavelength for *A. platensis* (618-626 nm) implies that the quantum yield was conserved in this case and then, that the $P/2e^-$ and the global stoichiometry are also conserved. Thus, even if the photosynthetic quotient was not determined experimentally in this work, it can be concluded that it is not likely to be affected when using optimal red light, enabling the calculation of the $O_2$ regenerating performances of any PBR if the air revitalization in closed life support systems was envisaged.

**ACKNOWLEDGMENT**

This work was supported by the European Space Agency (ESA/ESTEC) through the Biorat project. The authors also thank Pascal Lafon for his technical assistance.

## NOMENCLATURE

| | | |
|---|---|---|
| $\mathcal{A}$ | Local volumetric rate of radiant energy absorbed | [µmol.s$^{-1}$.m$^{-3}$] |
| $b$ | Back-scattered fraction for radiation | [dimensionless] |
| $C_X$ | Biomass concentration | [kg.m$^{-3}$ or g.L$^{-1}$] |
| $e_\lambda$ | Spectral efficiency | [dimensionless] |
| $Ea$ | Mass absorption coefficient | [m$^2$.kg$^{-1}$] |
| $E_{elec}$ | Luminous efficiency | [lm.W$_{elec}^{-1}$] |
| $E_{lum}$ | Radiant efficiency | [lm.W$_{rad}^{-1}$] |
| $Es$ | Mass scattering coefficient | [m$^2$.kg$^{-1}$] |
| $G$ | Local spherical irradiance | [µmol.s$^{-1}$.m$^{-2}$] |
| $I$ | Luminous intensity | [cd] |
| $J_i$ | Molar specific rate for species $i$ | [mol.kg$_X^{-1}$.s$^{-1}$] |
| $K$ | Half saturation constant for photosynthesis | [µmol.s$^{-1}$.m$^{-2}$] |
| $L$ | Total length of the photobioreactor | [m] |
| $L_l$ | Lightening length inside a rectangular photobioreactor | [m] |
| $M_X$ | C-molar mass for *Arthrospira platensis* | [kg$_X$.mol$_X^{-1}$] |
| $p(\theta,\theta')$ | Phase function for scattering | [dimensionless] |
| $p_\lambda^{CIE}$ | Normalized CIE photopic curve | [dimensionless] |
| $q$ | Photon flux density | [µmol.s$^{-1}$.m$^{-2}$] |
| $Q_P$ | Photosynthetic quotient ($Q_P = <r_{O_2}>/<r_{CO_2}>$) | [dimensionless] |
| $r$ | Optical reflection factor at a given interface | [dimensionless] |
| $r_X$ | Biomass volumetric growth rate (productivity) | [kg.m$^{-3}$.h$^{-1}$ or g.L$^{-1}$.h$^{-1}$] |
| $t$ | Time | [s or h] |
| V | Volume | [m$^3$ or L] |
| $V_l$ | Lightening volume inside the photobioreactor | [m$^3$ or L] |
| $w_\lambda$ | Spectral energetic emission spectrum of a source | [W$_{rad}$.m$^{-1}$] |
| $w_\lambda'$ | Spectral photonic emission spectrum of a source | [µmol.s$^{-1}$.m$^{-1}$] |
| $W_{elec}$ | Nominal electrical power for LED | [W$_{elec}$] |
| $W_{rad}$ | Radiant power | [W$_{rad}$] |
| $W_{photon}$ | Photon flow rate | [µmol$_{hv}$.s$^{-1}$] |
| $z$ | Length | [m] |

**Greek letters**

| | | |
|---|---|---|
| $\alpha$ | Linear scattering modulus | [dimensionless] |
| $\gamma$ | Fraction for working illuminated volume in the photobioreactor | [dimensionless] |
| $\Gamma$ | Quantum yield for LED | [$\mu mol_{h\nu}.s^{-1}.W_{elec}^{-1}$] |
| $\delta$ | Extinction coefficient | [$m^{-1}$] |
| $\eta$ | Energetic yield for LED | [dimensionless] |
| $\theta, \theta'$ | Incident and scattering angles | [rad] |
| $\kappa$ | Imaginary part of the complex refractive index for the micro-organism | [dimensionless] |
| $\lambda$ | Wavelength | [m] |
| $\Xi$ | Photonic conversion factor | [$\mu mol_{h\nu}.s^{-1}.W_{lum}^{-1}$] |
| $\rho$ | Energetic yield for photon conversion | [dimensionless] |
| $\rho_M$ | Maximum energetic yield for photon conversion | [dimensionless] |
| $\tau$ | Residence time | [h] |
| $\upsilon_{ij}$ | Stoichiometric coefficient | [dimensionless] |
| $\phi$ | Mass quantum yield for the Z-scheme of photosynthesis | [$kg_X.\mu mol_{h\nu}^{-1}$] |
| $\phi'$ | Mole quantum yield for the Z-scheme of photosynthesis | [$mol_X.\mu mol_{h\nu}^{-1}$] |
| $\Omega$ | Solid angle | [rad] |

**Subscripts**

| | |
|---|---|
| $\cap$ | Relative to hemispherical incident radiation onto the photobioreactor |
| $\lambda$ | Relative to a spectral quantity for the wavelength $\lambda$ |
| $\Delta\lambda$ | Relative to a narrow interval of wavelength |
| min | Minimum value for the spectral range of wavelength |
| max | Maximum value for the spectral range of wavelength |

**Other**

| | |
|---|---|
| $\overline{X}$ | Spectral averaging |
| $<X> = \dfrac{1}{V}\iiint_V X\, dV$ | Spatial averaging |

**APPENDIX**

**Calculation of Energetic and Quantum Yields of LED**

Generally speaking, the energetic yield of a light source $\eta$ ($W_{rad}$ / $W_{elec}$) is defined as the ratio of the radiant light power output over the electrical power input. It can be easily converted into in quantum yield $\Gamma$ (µmol photon per second and per electrical watt) when the conversion factor $\Xi$ (µmol photon per second and per radiant watt) of the source is known.

First, the conversion factor is easily derived from the knowledge of the spectral energetic emission spectrum of the source $w_\lambda$ ($W_{rad}.m^{-1}$ which is formally the radiant power transported by photons of wavelength $\lambda$) and from its conversion into photonic unit spectrum $w'_\lambda$ (µmol photon.$s^{-1}.m^{-1}$ which is formally the flow rate of photons of wavelength $\lambda$). The radiant light power is given by:

$$W_{rad} = \int_{\lambda_{min}}^{\lambda_{max}} w_\lambda d\lambda \quad \text{(radiant watt)}$$

The photon flow rate is given by:

$$W_{photon} = \int_{\lambda_{min}}^{\lambda_{max}} w'_\lambda d\lambda \quad (\mu\text{mol photon.s}^{-1})$$

The relationship between $w_\lambda$ and $w'_\lambda$ results from quantum mechanisms:

$$w_\lambda = w'_\lambda\, 6.02 \times 10^{17} \frac{hc}{\lambda} = 1.1974 \times 10^{-7} \frac{w'_\lambda}{\lambda} \quad (A1)$$

The conversion factor $\Xi = \dfrac{W_{photon}}{W_{rad}}$ (µmol photon.s$^{-1}$/radiant watt) is then:

$$\Xi = 8.351\times 10^6 \frac{\int_{\lambda_{min}}^{\lambda_{max}} w_\lambda \lambda\, d\lambda}{\int_{\lambda_{min}}^{\lambda_{max}} w_\lambda\, d\lambda} = 8.351\times 10^6\, \lambda_{m,w} \quad (A2)$$

where $\lambda_{m,w}$ is the average energetic wavelength defined by: $\lambda_{m,w} = \dfrac{\int_{\lambda_{min}}^{\lambda_{max}} w_\lambda \lambda\, d\lambda}{\int_{\lambda_{min}}^{\lambda_{max}} w_\lambda\, d\lambda}$ and is characteristic of the emission spectrum of any given source.

If we assume a solar spectrum or a typical white fluorescent lamp with a color temperature around 5500 K, this average wavelength is roughly 550 nm leading to the well-known conversion factor in the PAR (photosynthetically active radiation) with $\Xi = 4.6$ µmol photon.s$^{-1}$ per radiant watt.

Secondly, the energetic yield $\eta$ is calculated by introducing the photometry units used by the manufacturers of artificial light sources. Generally, the key technical data available is the luminous efficiency $E_{elec}$ (in lumens per electrical watt). In order to define the energetic yield $\eta$ (radiant watt per electrical watt) it is then convenient to define the radiant efficiency $E_{lum}$ in lumens per radiant watt.

This quantity is also determined from the knowledge of the energetic emission spectrum of the source $w_\lambda$, considering the CIE normalized photopic response of human eye $p_\lambda^{CIE}$ (see Figure A1) and the definition of the candela SI unit, from:

$$E_{lum} = \frac{683 \int_{\lambda_{min}}^{\lambda_{max}} w_\lambda p_\lambda^{CIE} \, d\lambda}{\int_{\lambda_{min}}^{\lambda_{max}} w_\lambda \, d\lambda} \quad \text{(lumen per radiant watt)} \quad (A3)$$

The energetic yield, as previously defined, is then:

$$\eta = \frac{W_{rad}}{W_{elec}} = \frac{E_{elec}}{E_{lum}} \quad (A4)$$

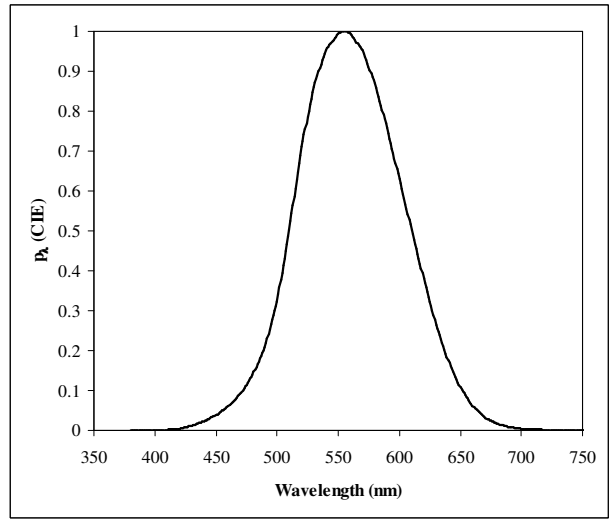

**Figure A1**. Normalized and standard CIE photopic curve $p_\lambda^{CIE}$. The maximum value ($p_\lambda^{CIE} = 1$) is reached for the green wavelength at 555 nm.

On the contrary, for LED, the former quantity $E_{elec}$ is generally not directly available from manufacturers, but may be calculated from basic data as luminous intensity $I$ (in Candela), angle of emission $\theta$, and nominal electrical power consumption $W_{elec}$ by:

$$E_{elec} = \frac{I \, d\Omega}{W_{elec}} = \frac{2\pi I (1-\cos\theta)}{W_{elec}} \quad (A5)$$

enabling the calculation of the energetic yield with equation (4), and then the quantum yield, applying:

$$\Gamma = \Xi \, \eta \quad (A6)$$

**Choice of the LED Panels Composition**

First the control panel of LED has been defined in such a way that the two photosystems of *Arthrospira* be excited in series in the red domain (PSII) and in the blue domain (PSI) with a P/2e$^-$ = 1.275 ratio corresponding to the established stoichiometry for low incident light fluxes used in this study (see the modeling part in the text). The green light has been removed from the control panel because of its poor absorption by pigments in the PBR. From the technical data on the blue and red LED used (see the Materials and Methods part), and combining their emission spectrum with equations (A1-A6), the following energetic and quantum yields are calculated leading to:

- Blue LED (Toyoda Gosei E1L51-3B0A, 470 nm): $\eta$ = 0.057 radiant watt per electrical watt – $\Gamma$ = 0.22 µmol photon per second and per electrical watt.

- Red LED (Fairchild MV8014, 640 nm): $\eta$ = 0.041 radiant watt per electrical watt – $\Gamma$ = 0.22 µmol photon per second per electrical watt.

This result explains why these LED were chosen to give the same quantum yield $\Gamma$, enabling to simply obtain the target value for the ratio P/2e$^-$ = 1.275 just by building a panel with 1.275 blue LED for 1 red LED. Such a control panel delivers for its nominal electrical power an incident light flux $q_\cap$ = 33 $\mu mol_{hv}.m^{-2}.s^{-1}$ (corresponding effectively to 18.5 $\mu mol_{hv}.m^{-2}.s^{-1}$ for blue LED and 14.5 $\mu mol_{hv}.m^{-2}.s^{-1}$ for red LED as it should be).

Secondly, the high efficiency red panel has been conceived to work in the optimal range of wavelength defined from the spectral efficiency $e_{\Delta\lambda} \cong 1$ (see Figure 3 in text). The same calculations of yields as previously done for these LED lead to:

- High efficiency red LED (Agilent HLMP-EG08-WZ000, 626 nm: $\eta = 0.171$ radiant watt per electrical watt – $\Gamma = 0.90$ $\mu$mol photon per second and per electrical watt.

Such a high efficiency panel delivers, for the same nominal electrical power as the control panel, an incident light flux $q_\cap = 135$ $\mu mol_{hv}.m^{-2}.s^{-1}$. This experimental value, together with the previous value of the control panel, clearly validates the theoretical approach in the first part of this appendix, considering the yields of LED calculation. They verify indeed:

$$\frac{\Gamma \text{ (high efficiency red LED)}}{\Gamma \text{ (control, mixed blue/red LED)}} \cong \frac{0.9}{0.22} \cong \frac{135}{33} \cong 4.1$$

as it should be, demonstrating that using a high efficiency red LED panel can save in our case a factor 4 for electrical power consumption with the same incident light flux $q_\cap$.